\documentclass[aps,prd,twocolumn,nofootinbib,superscriptaddress]{revtex4-2}

\usepackage{graphicx}
\usepackage{epstopdf}
\usepackage{natbib}
\usepackage{latexsym}
\usepackage{amssymb}
\usepackage{amsmath}
\usepackage{color}
\usepackage{mathrsfs}
\usepackage{xparse}
\usepackage{bbding}
\usepackage{enumitem}
\usepackage{pifont}
\usepackage[center]{subfigure}
\usepackage[
            pdfstartview=FitH,
            bookmarksnumbered=true,
            bookmarksopen=true,
            colorlinks,
            linkcolor=blue,
            anchorcolor=green,
            citecolor=blue
            ]{hyperref}

            \usepackage{setspace}
\usepackage{amssymb}
\usepackage{graphicx,wrapfig}
\usepackage{enumerate}
\usepackage{color}
\usepackage{mathrsfs}
\usepackage{subfigure}
\usepackage{comment}

\definecolor{mg}{rgb}{0.0, 0.5, 0.0}

\allowdisplaybreaks[4]
\begin{document}

  \renewcommand\arraystretch{2}
 \newcommand{\bq}{\begin{equation}}
 \newcommand{\eq}{\end{equation}}
 \newcommand{\bqn}{\begin{eqnarray}}
 \newcommand{\eqn}{\end{eqnarray}}
 \newcommand{\nb}{\nonumber}
 \newcommand{\lb}{\label}

\newcommand{\La}{\Lambda}
\newcommand{\va}{\scriptscriptstyle}
\newcommand{\be}{\nopagebreak[3]\begin{equation}}
\newcommand{\ee}{\end{equation}}

\newcommand{\ba}{\nopagebreak[3]\begin{eqnarray}}
\newcommand{\ea}{\end{eqnarray}}

\newcommand{\la}{\label}
\newcommand{\n}{\nonumber}
\newcommand{\su}{\mathfrak{su}}
\newcommand{\SU}{\mathrm{SU}}
\newcommand{\U}{\mathrm{U}}

\def\be{\nopagebreak[3]\begin{equation}}
\def\ee{\end{equation}}
\def\ba{\nopagebreak[3]\begin{eqnarray}}
\def\ea{\end{eqnarray}}
\newcommand{\f}{\frac}
\def\rmd{\rm d}
\def\lp{\ell_{\rm Pl}}
\def\d{{\rm d}}
\def\fe{\mathring{e}^{\,i}_a}
\def\fw{\mathring{\omega}^{\,a}_i}
\def\fq{\mathring{q}_{ab}}
\def\t{\tilde}

\def\db{\delta_b}
\def\dc{\delta_c}
\def\T{\mathcal{T}}
\def\GammaE{\Gamma_{\rm ext}}
\def\GammaEb{\bar\Gamma_{\rm ext}}
\def\GammaEh{\hat\Gamma_{\rm ext}}
\def\Hee{H_{\rm eff}^{\rm ext}}
\def\H{\mathcal{H}}

\newcommand{\R}{\mathbb{R}}

 \newcommand{\cb}{\color{blue}}
    \newcommand{\cc}{\color{cyan}}
        \newcommand{\cm}{\color{magenta}}
\newcommand{\rc}{\rho^{\scriptscriptstyle{\mathrm{I}}}_c}
\newcommand{\rd}{\rho^{\scriptscriptstyle{\mathrm{II}}}_c}
\NewDocumentCommand{\evalat}{sO{\big}mm}{%
  \IfBooleanTF{#1}
   {\mleft. #3 \mright|_{#4}}
   {#3#2|_{#4}}%
}
\newcommand{\PRL}{Phys. Rev. Lett.}
\newcommand{\PL}{Phys. Lett.}
\newcommand{\PR}{Phys. Rev.}
\newcommand{\CQG}{Class. Quantum Grav.}



\title{On the improved dynamics approach in loop quantum black holes}

\author{Hongchao Zhang}
\email{zhanghongchao852@live.com}
\affiliation{Institute for Theoretical Physics \& Cosmology, Zhejiang University of Technology, Hangzhou, 310023, China}
\affiliation{United Center for Gravitational Wave Physics (UCGWP),  Zhejiang University of Technology, Hangzhou, 310023, China}

\author{Wen-Cong Gan}
\email{Wen-cong$\_$Gan1@baylor.edu}
\affiliation{College of Physics and Communication Electronics, Jiangxi Normal University, Nanchang 330022, China}
\affiliation{GCAP-CASPER, Physics Department, Baylor University, Waco, Texas 76798-7316, USA}
\author{Yungui Gong}
\email{yggong@hust.edu.cn}
\affiliation{School of Physics, Huazhong University of Science and Technology, Wuhan, Hubei 430074, China}
\affiliation{Department of Physics, School of Physical Science and Technology, Ningbo University, Ningbo, Zhejiang 315211, China}

\author{Anzhong Wang}
\email{Corresponding author. anzhong$\_$wang@baylor.edu}
\affiliation{GCAP-CASPER, Physics Department, Baylor University, Waco, Texas 76798-7316, USA}

\date{\today}

\begin{abstract}

In this paper, we consider the  B\"ohmer-Vandersloot  (BV) model of loop quantum black holes obtained from  the improved dynamics approach.
We adopt the Saini-Singh gauge, in which it was found analytically that the BV spacetime is geodesically complete.  We  show that black/white
hole horizons do not exist  in this geodesically complete spacetime. Instead, there exists only an infinite number of transition surfaces, which always separate
trapped regions from anti-trapped ones. Comments on the improved dynamics approach adopted in other models of loop quantum black holes are also given.

\end{abstract}

\maketitle

\section{Introduction}
\label{sec:Intro}

In Einstein's general relativity (GR), two   different kinds of  spacetime singularities appear, one is the  big-bang singularity of our universe, and the other is the internal singularity of a black hole.  It is commonly understood that the spacetime curvatures become Planckian when very closed to these singularities, and GR ceases to be valid, as quantum gravitational effects  in such small scales become important and must be taken into account. It is our cherished hope that these singularities will be smoothed out  after  such quantum effects are taken into account.

In the past two decades, it has been shown that this is indeed the case for the big-bang singularity in the framework of  loop quantum cosmology (LQC) \cite{Bojowald:2001xe,Ashtekar:2006wn,Ashtekar:2011ni}. LQC is constructed by applying loop quantum gravity (LQG)  techniques to cosmological models within the superminispace approach \cite{Ashtekar:2003hd}, and the resulting quantum corrections to classical geometry can be effectively described by semiclassical effective Hamiltonian that incorporate the leading-order quantum geometric effects \cite{Taveras:2008ke}. The effective model works very well in comparison with the full quantum dynamics of LQC even in the deep quantum regime \cite{Ashtekar:2011ni}, especially for the states that are sharply peaked on a classical trajectory at late times \cite{Kaminski:2019qjn}.
LQC can resolve the big-bang singularity precisely because of the fundamental result of LQG:  {\em quantum gravity effects always lead  the area operator to have a non-zero minimal area gap} \cite{Thiemann:2007pyv}. It is this non-zero area gap that causes strong repulsive effects in the dynamics when the spacetime curvature reaches the Plank scale and the big-bang singularity is replaced by a quantum bounce \cite{Singh:2009mz}.

The semiclassical effective Hamiltonian can be obtained from the classical one simply by the replacement
\bqn
\lb{eq1.1}
c \rightarrow \frac{\sin\left(\mu c\right)}{\mu},
\eqn
where $c$ denotes the moment conjugate of the area operator $p$ ($\propto a^2$, where $a$ is the expansion factor of the universe), and $\mu$ is called the polymerization parameter. Clearly, when $\mu \rightarrow 0$, the classical limit is obtained, while when $\mu \gg 0$, the quantum gravitational effects become large, whereby a mechanism for resolving the big-bang singularity is provided.
In LQC, there exist two different quantization schemes, the so-called  $\mu_o$ and $\bar\mu$ schemes, which give different representations of quantum Hamiltonian constraints and lead to different effective dynamics \cite{Ashtekar:2011ni}. The fundamental difference of these two approaches rises in the implementation of the minimal area gap mentioned above.
In the $\mu_o$ scheme, each holonomy $h_k^{(\mu)}$ is considered as an eigenstate of the area operator, associated with the face of the elementary cell orthogonal to the $k$-th direction. The parameter $\mu$ is fixed by {requiring  the corresponding eigenvalue be the minimal area gap.} As a result, $\mu$ is a constant in this approach \cite{Ashtekar:2003hd}
\bqn
\lb{eq1.2}
\mu = {\text{Constant, say, }} \;\mu_o.
\eqn
  However, it has been shown \cite{Corichi:2009pp} that this quantization does not have a proper semiclassical limit, and suffers from the dependence on the  length of the fiducial cell. It also lacks of consistent identified curvature scales. On the other hand, in the $\bar\mu$ scheme \cite{Ashtekar:2006wn}, the quantization of areas is referred to the physical geometries, and when shrinking a loop until the minimal area enclosed by it, one should use the physical geometry. Since the latter depends on the phase space variables, now when calculating the holonomy $h_k^{(\mu)}$, one finds that the parameter $\mu$ depends on the phase space variable $p$ \cite{Ashtekar:2006wn}
\bqn
\lb{eq1.3}
\mu = \bar\mu \equiv \frac{\Delta}{|p|},
\eqn
where $\Delta \equiv \left(4\sqrt{3} \pi \gamma\right) \ell^2_{pl}$, with $\gamma$ being the Barbero-Immirzi parameter and $ \ell_{pl}$ the Planck length. When the expansion factor is very large we have
$|p| \gg \Delta$, so that $\bar\mu \rightarrow 0$, and the quantum effects are expected to be very small. However, near the singular point $|p| \simeq 0$, we have $\bar\mu \gg 1$, so that the quantum effects
are expected to be very large so that the big-bang singularity
used to appear at $|p|  = 0$ now is replaced by a quantum bounce.
In the literature, this improved dynamical approach is often referred to as the $\bar\mu$ scheme, and has been shown to be the only scheme discovered so far  that overcomes the  limitations of the $\mu_o$ scheme and is consistent with observations \cite{Ashtekar:2011ni}.

In parallel to the studies of LQC, loop quantum black holes (LQBHs) have been also intensively studied in the past decade or so (See, for example, \cite{Olmedo:2016ddn,Ashtekar:2018cay,Ashtekar:2020ifw,Gambini:2022hxr,Ashtekar:2023cod,Lewandowski:2022zce} and references therein). In particular,  since the spacetime of the  Schwarzschild black hole interior is homogeneous
 and the metric is only time-dependent, so it can  be treated as the Kantowski-Sachs spacetime
\bq
\lb{eq1.4}
 \text{d}s^2 = - N^2(T) \text{d}T^2 + \frac{p_b^2(T)}{L_o^2 \left|p_c(T)\right|} \text{d}x^2 + \left|p_c(T)\right| \text{d}\Omega^2,
 \eq
 where  $L_o$ denotes the length of the fiducial cell in the $x$-direction, and $\text{d}\Omega^2 \equiv \text{d}\theta^2 + \sin^2\theta \text{d}\phi^2$. Then, some  LQC techniques can be borrowed  to study the black hole interiors directly.
 In particular,   LQBHs were  initially studied within the  $\mu_o$ scheme \cite{Modesto:2004wm,Ashtekar:2005qt,Modesto:2005zm}. However, this LQBH model also suffers from similar limitations as the $\mu_o$ scheme in LQC \cite{Corichi:2015xia,Olmedo:2017lvt,Ashtekar:2018cay}. Soon the $\bar \mu$ scheme was applied to the Schwarzschild black hole interior by B\"ohmer and Vandersloot (BV) \cite{Boehmer:2007ket} with the replacements
 \bq
\lb{eq1.5}
b \rightarrow \frac{\sin(\delta_b b)}{\delta_b}, \quad c \rightarrow \frac{\sin(\delta_c c)}{\delta_c},
\eq
in the classical  Hamiltonian, where $b$ and $c$ are the moment conjugates of $p_b$ and $p_c$, with $\{c,p_c \}=2G \gamma$, $\{b,p_b \}=G \gamma$, and $\delta_b$ and $\delta_c$ are the corresponding
two polymerization parameters, given by \cite{Boehmer:2007ket}
 \bqn
\lb{eq1.6}
\delta_b = \sqrt{\frac{\Delta}{|p_c|}}, \quad
\delta_c = \frac{\sqrt{\Delta {|p_c|}}}{p_b}.
\eqn
To understand the quantum effects, let us first note that in the interior of the Schwarzschild black hole we have \cite{Gan:2022oiy}
\bqn
\lb{eq1.7}
p_b^{\text{GR}}  = \text{e}^{T}\sqrt{2m \text{e}^{-T} - 1},\;\;\;
p_c^{\text{GR}} =  \text{e}^{2T},
\eqn
for which the
black hole singularity is located at $T = -\infty$, while its horizon is located at $T = T^{\text{GR}}_H \equiv \ln(2m)$. Thus, near the singular point we have $\delta_b \propto \text{e}^{-T} \rightarrow \infty$, although   $\delta_c \propto \text{e}^{T/2} \rightarrow 0$.
Then, we expect that the quantum effects become so large that the curvature singularity is smoothed out and finally replaced by a regular transition surface  \cite{Boehmer:2007ket}. On the other hand, near the black hole horizon, we have  $p_c^{\text{GR}} \simeq 4m^2$
and $p_b^{\text{GR}}  \simeq 0$, so that $\delta_c \rightarrow  \infty$ (although  now $\delta_b$ remains finite). Then, we expect that {\em there are large departures from the classical theory very near the classical black hole horizon even for massive black holes, for which the curvatures at the horizon become very low} \cite{Boehmer:2007ket,Corichi:2015xia,Olmedo:2017lvt,Ashtekar:2018cay}. As a matter of fact, recently we found that the effects are so large that
{\em black/white horizons never exist in the BV model} \cite{Gan:2022oiy}.

It should be noted that in \cite{Gan:2022oiy} the lapse function was chosen as $N = {\gamma\delta_b \sqrt{|p_c|}}/{\sin(\delta_b b)}$, in which the coordinate $T$ does not represent the cosmic time. Then, one may wonder if $T$ covers the whole
spacetime of the BV model. On the other hand,  in \cite{Saini:2016vgo} the proof that the BV model is geodesically complete was carried out in the cosmic time coordinate, in which the lapse function was set to one. In this paper, we shall adopt the Saini-Singh (SS) gauge, $N = 1$, and show that  indeed black/white horizons never exist in the BV model, as expected from what we obtained in \cite{Gan:2022oiy}, since the physics should not depend on the choice of the gauge.

The rest of the paper is organized as follows: In the next section, Sec. II, we first re-derive the corresponding field equations in the SS gauge, and correct typos existing in the literature. Then, we re-confirm the result obtained in
\cite{Saini:2016vgo} that the BV spacetime is geodesically complete even without matter. After that, from the definitions of black/white hole horizons we show explicitly that they do not exist in the BV model. Instead, there exist infinite
regular transition surfaces that always separate a trapped region from an anti-trapped one. Finally, in Sec. III, we present our main conclusions and provide comments on other models of LQBHs, adopting the $\bar\mu$ scheme. In particular, the models studied recently by Han and Liu \cite{Han:2020uhb,Han:2022rsx} are absent of the above pathology.

\section{ B\"ohmer-Vandersloot Model with  Saini-Singh Gauge}\lb{sec-bv}

 To show our above claim, we first note that  the Kantowski-Sachs metric (\ref{eq1.4}) is invariant under the gauge transformations
 \bq
 \lb{eq2.3}
 T = f(\tau), \quad x = \alpha \hat x + x_o,
 \eq
via the redefinitions of the lapse function and the length of the fiducial cell,
 \bq
 \lb{eq2.4}
 \hat N = Nf_{,\tau}, \quad \hat{L}_o = \frac{L_o}{\alpha},
 \eq
 where $f(\tau)$ is an arbitrary function of $\tau$, and $\alpha$ and $x_o$ are arbitrary but real constants. Using the above freedom, we can always choose the SS gauge  \cite{Gan:2022oiy}
 \bq
 \lb{eq2.5}
 N = 1.
 \eq
For this particular choice of the gauge, we denote the timelike coordinate $T$ by $\tau$. Then, the corresponding effective BV Hamiltonian  reads  \cite{Saini:2016vgo}
 \begin{widetext}
\bqn
\lb{eq2.6}
H^{\text{eff}}[N=1] &=& - \frac{p_b \sqrt{p_c}}{2\gamma^2 G \Delta}\Bigg[2 \sin(\delta_b b) \sin(\delta_c c)  + \sin^2(\delta_b b) + \frac{\gamma^2\Delta}{p_c}\Bigg],
\eqn
from which we find the equations of motion (EoMs) are given by
\bqn\lb{eq2.7}
\dot b&=& G \gamma  \frac{\partial H^{\text{eff}}}{\partial p_b}=\frac{c p_c}{\gamma \sqrt{\Delta} p_b} \sin(\delta_b b) \cos(\delta_c c),\nb\\
\\
\lb{eq2.8}
\dot c&=&2G \gamma  \frac{\partial H^{\text{eff}}}{\partial p_c}=-\frac{c}{\gamma \sqrt{\Delta}} \sin(\delta_b b) \cos(\delta_c c)+\frac{b p_b}{\gamma \sqrt{\Delta} p_c}\cos(\delta_b b) \left(\sin(\delta_b b)+\sin(\delta_c c)\right)+\frac{\gamma p_b}{p_c^{3/2}},\\
\lb{eq2.9}
\dot p_c &=&-2G \gamma  \frac{\partial H^{\text{eff}}}{\partial c} = \frac{2 p_c}{\gamma \sqrt{\Delta}} \sin(\delta_b b) \cos(\delta_c c), \\
\lb{eq2.10}
\dot p_b &=&-G \gamma  \frac{\partial H^{\text{eff}}}{\partial b} = \frac{p_b}{\gamma \sqrt{\Delta}} \cos(\delta_b b) \left(\sin(\delta_b b)+\sin(\delta_c c)\right).
\eqn
Note that in writing down the above equations, we had used the Hamiltonian constraint $H^{\text{eff}} \approx 0$. It should be also noted that    there exists a typo in the EoM of $c(\tau)$ given in \cite{Saini:2016vgo},
 where the last term should be
 ${\gamma p_b}/{p_c^{3/2}}$, instead of ${\gamma p_b}/{{p_c^{1/2}}}$ [cf. Eq.(3.6) of \cite{Saini:2016vgo} and recall that now we consider the vacuum case $\rho = 0$].
  From Eqs.(\ref{eq2.9}) and (\ref{eq2.10}), we find that
 \bqn
 \lb{eq2.11}
  p_c &=& p_c^{(0)}\exp\left\{\frac{2}{\gamma \sqrt{\Delta}}\int_{\tau_0}^{\tau}{ \sin(\delta_b b) \cos(\delta_c c)d\tau'}\right\},\nb\\
   p_b &=& p_b^{(0)}\exp\left\{\frac{1}{\gamma \sqrt{\Delta}}\int_{\tau_0}^{\tau}{ \cos(\delta_b b) \Big[\sin(\delta_b b)+\sin(\delta_c c)\Big]d\tau'}\right\},
  \eqn
   \end{widetext}
  where $p_c^{(0)}$ and $p_b^{(0)}$ are two integration constants.
  Note that the intergrades of both $p_c$ and $p_b$ are less or maximally equal to two at any given moment $\tau$, so
  we must have
  \bqn
 \lb{eq2.12}
  0 < p_b, \; p_c <  \infty,
   \eqn
  within  any given finite time $\tau$  \cite{Saini:2016vgo}. As a result, the  range of $\tau \in(-\infty, \infty)$ cover the whole spacetime, and the corresponding BV universe is geodesically complete in the ($\tau, x, \theta, \phi)$-coordinates.
  In particular, $p_c(\tau = \infty) = \infty$ corresponds to the spacetime casual boundaries, and no extensions beyond it are needed.

   In Figs. \ref{fig1} and \ref{fig2} we plot the
  four physical variables $\left(b, c, p_b, p_c\right)$ for $-3<\tau < 415$ and  $m/\ell_{pl}=1$ with
  the  initial time being chosen at $\tau_i = -2.15018$, and the initial data $\left.\left(b, c, p_b, p_c\right)\right|_{\tau_i}$ as those given in \cite{Gan:2022oiy}, in order to compare the results obtained in this two papers,
  where
  $\tau_{\cal{T}} \simeq 1.11068$ corresponds to the location of the first transition surface of the BV model, and $\tau^{\text{GR}}_H =0$  to the location of the classical Schwarzschild black hole horizon.
  In particular, we find that $p_c$ and $p_b$ are indeed finite and non-zero. This is true also for $b$, $c$, $g_{xx}$ and the Kretchmann scalar $K \equiv R_{\alpha\beta\gamma\delta}R^{\alpha\beta\gamma\delta}$.
  When $\tau \gg \tau_i$, we find that $p_c(\tau)$ is exponentially increasing. To monitor the numerical errors, we also plot out  $N^2$ and $\left|H^{\text{eff}}\right|$, from which one can see that $\left|H^{\text{eff}}\right| \le 6\times 10^{-12}$
  over the whole range of $\tau$.  To compare our results with those given in  \cite{Boehmer:2007ket},
  in Fig. \ref{fig3} we also plot out the corresponding physical quantities  for $\tau <  \tau_{\cal{T}}$. From it, it can be seen that our results match very well with those presented in  \cite{Boehmer:2007ket}.
  We also consider other choices of the mass parameter, and similar results are obtained.
  In review of all the above, one can see that our numerical code is quite trustable.

\begin{figure*}[htbp]
	\resizebox{\linewidth}{!}{\begin{tabular}{cccc}
			\includegraphics[height=4cm]{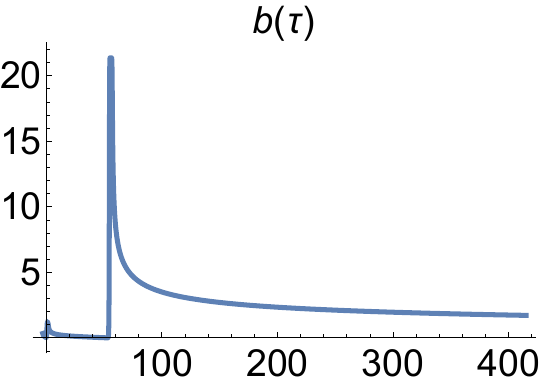}&
			\includegraphics[height=4cm]{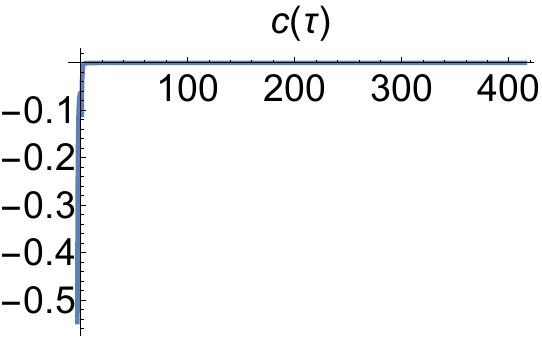}&
			\includegraphics[height=4cm]{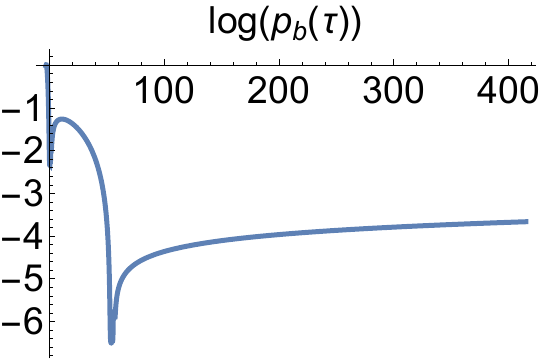}&
			\includegraphics[height=4cm]{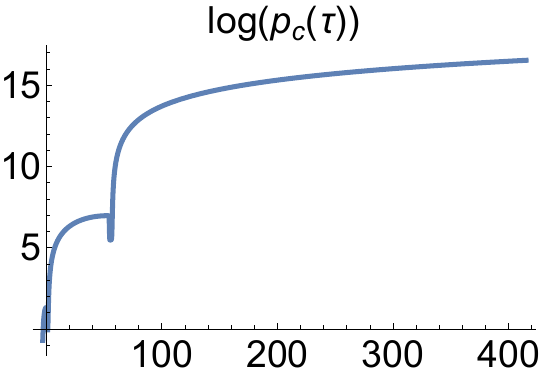}\\
		(a) & (b)&	(c) & (d) \\[6pt]
			\includegraphics[height=4cm]{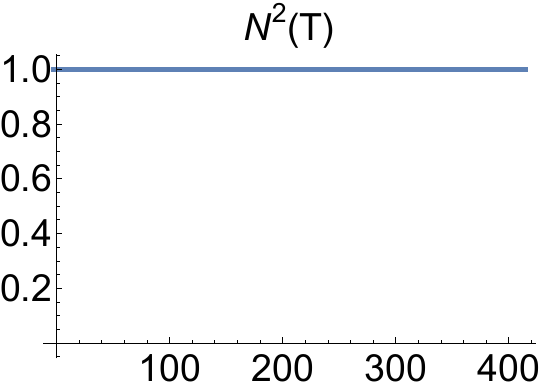}&
			\includegraphics[height=4cm]{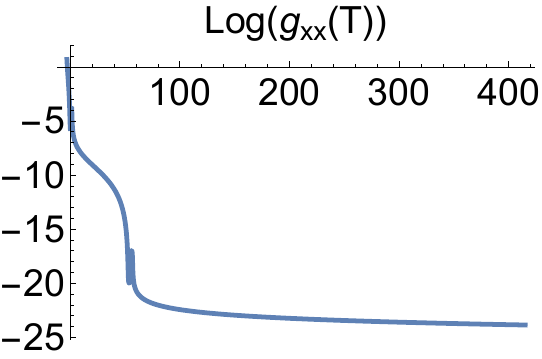}&
			\includegraphics[height=4cm]{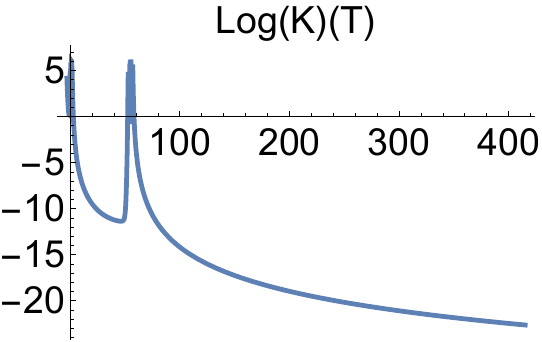}&
			\includegraphics[height=4cm]{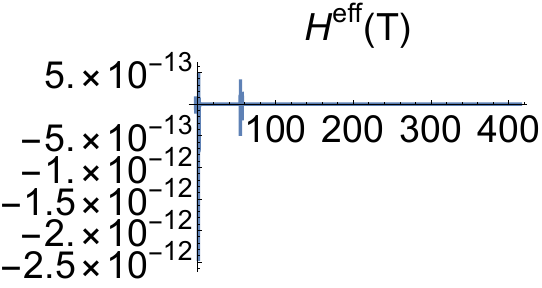}\\
				(e) & (f) &(g) & (h)  \\[6pt]
	\end{tabular}}
	\caption{Plots of  the four physical variables $\left(b, c, p_b, p_c\right)$ for $-3<\tau < 415$ and  $m/\ell_{pl}=1$, for which we have
		$\tau_{\cal{T}} \simeq 1.11068$ and $\tau^{\text{GR}}_H =0$. The initial time is chosen at $\tau_i = -2.15018$.}
	\lb{fig1}
\end{figure*}

\begin{figure*}[htbp]
 \resizebox{\linewidth}{!}{\begin{tabular}{cccc}
 \includegraphics[height=4cm]{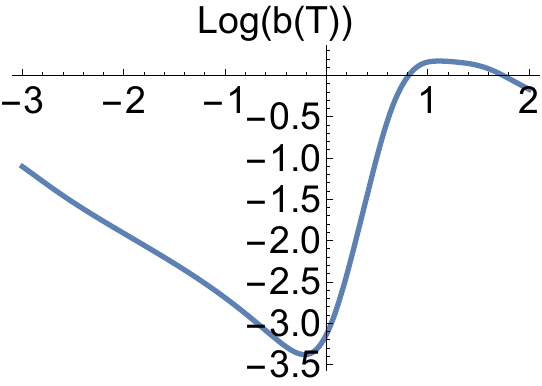}&
\includegraphics[height=4cm]{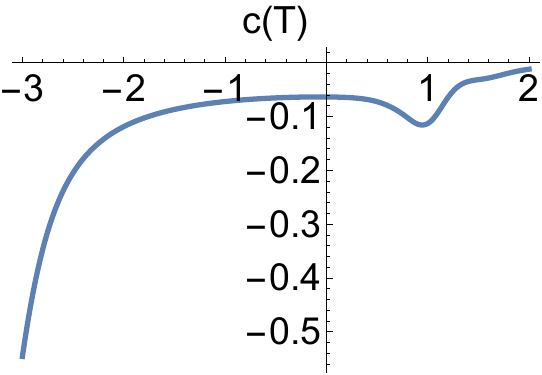}&
\includegraphics[height=4cm]{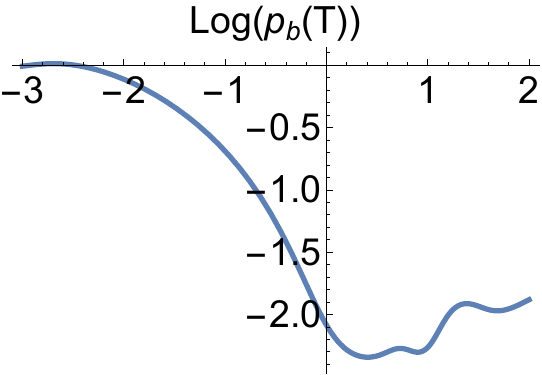}&
\includegraphics[height=4cm]{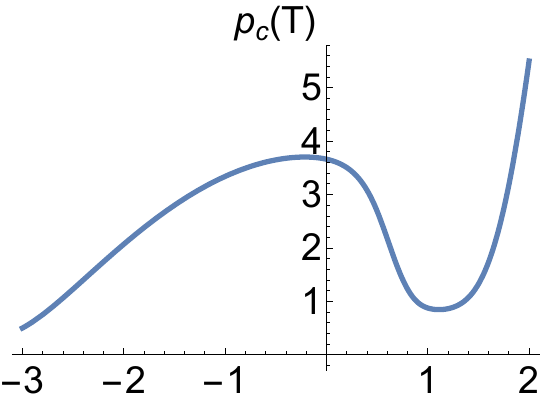}\\
 (a) & (b) &(c) & (d) \\[6pt]
\\
	\includegraphics[height=4cm]{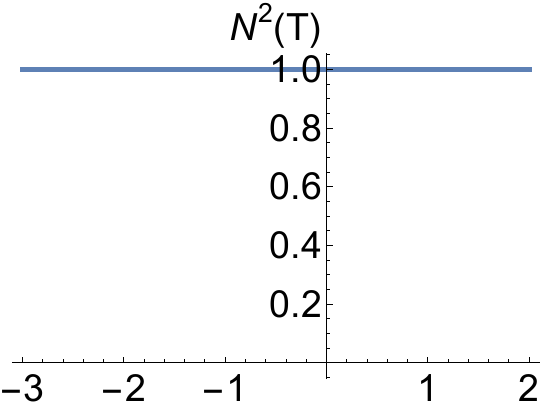}&
\includegraphics[height=4cm]{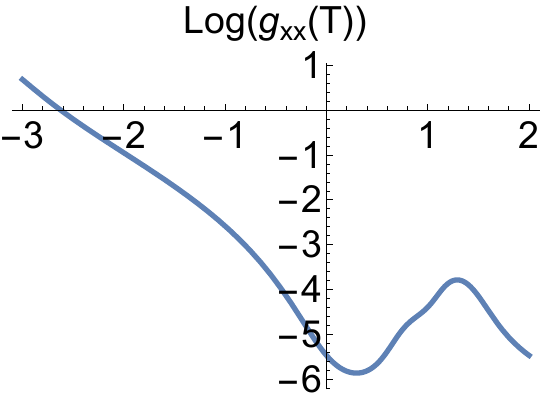}&
\includegraphics[height=4cm]{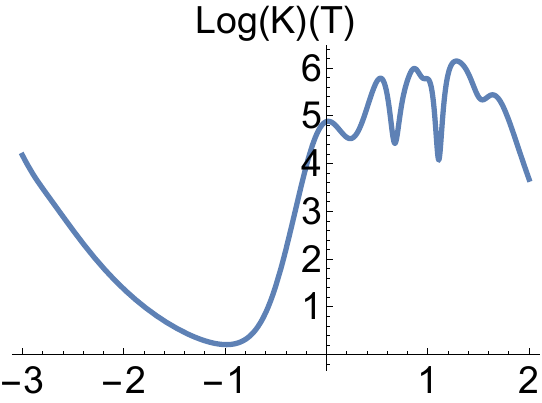}&
\includegraphics[height=4cm]{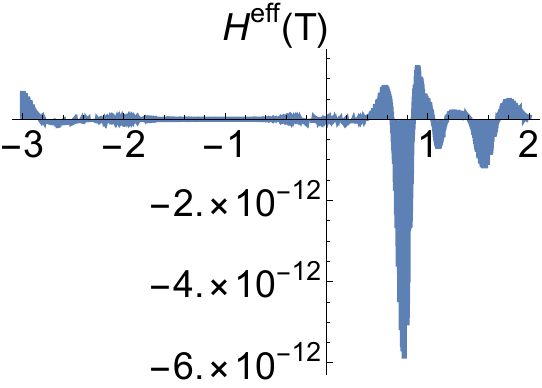}\\
(e) & (f) &	(g) & (h)  \\[6pt]
	\end{tabular}}
\caption{Plots of  $\left(b, c, p_b, p_c\right)$  for $-3<\tau < 2$.
The same initial time and conditions are chosen as those of Fig. \ref{fig1}.}
\lb{fig2}
\end{figure*}

\begin{figure*}[htbp]
 \resizebox{\linewidth}{!}{\begin{tabular}{cccc}
 \includegraphics[height=4.cm]{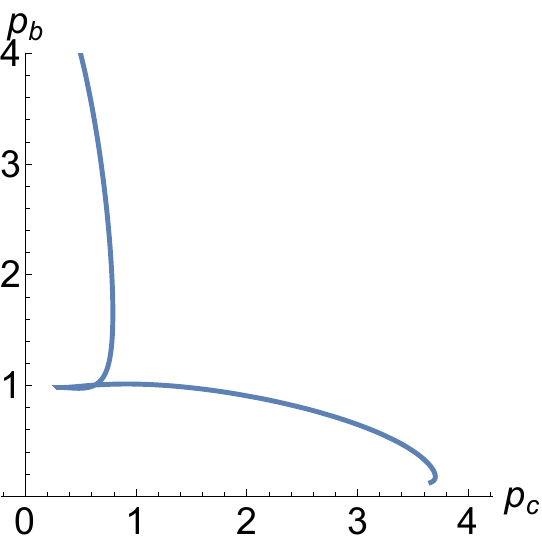}&
\includegraphics[height=4.cm]{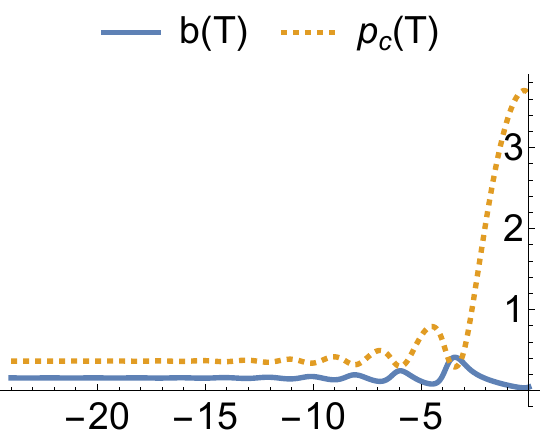}&
\includegraphics[height=4cm]{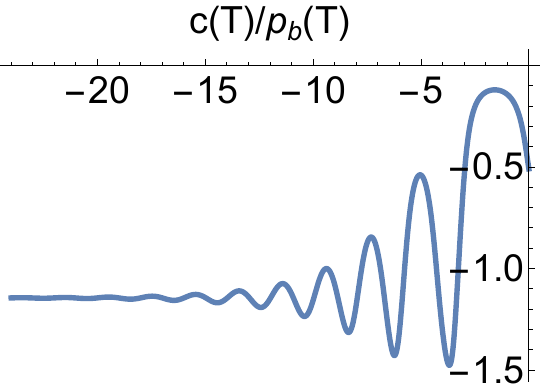} &
\includegraphics[height=4.cm]{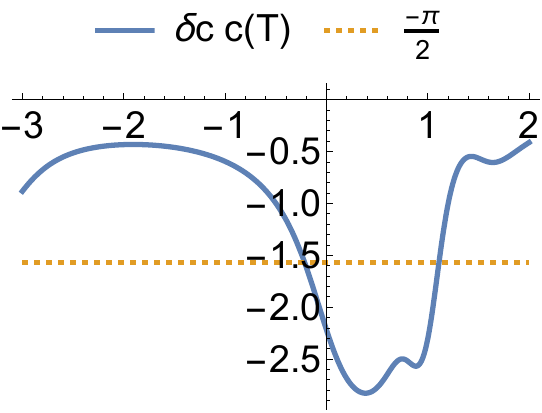}\\
 (a) & (b) &(c) & (d) \\[6pt]
	\end{tabular}}
\caption{Plots of  $\left(b, c, p_b, p_c\right)$  for $-24<\tau < 0$ and  $\delta_c c(\tau)$ for $-3<\tau < 2$ with $m/\ell_{pl}=1$.   The same initial time and conditions are chosen as those of Fig. \ref{fig1}.}
\lb{fig3}
\end{figure*}

  In the following, we shall show that in the geodesically complete BV model, black/white horizons never exist. Instead, only regular  transition surfaces exist. To show these claims, let us
first  introduce the unit vectors, $u_{\mu} \equiv  \delta^\tau_{\mu}$ and $s_{\mu} \equiv \sqrt{g_{xx}}\delta^x_{\mu}$. Then, we  construct  two null vectors $\ell_{\mu}^{\pm} = \left(u_{\mu} \pm s_{\mu} \right)/\sqrt{2}$, which define, respectively, the in-going and out-going radially-moving null geodesics. Then, the expansions of them are defined by \cite{Ashtekar:2018cay}
\bq
\lb{eq2.13}
\Theta_{\pm} \equiv m^{\mu\nu}\nabla_{\mu}\ell_{\nu}^{\pm} = - \frac{p_{c, \tau}}{\sqrt{2}  p_c},
\eq
 where $m_{\mu\nu} \equiv g_{\mu\nu} + u_{\mu}u_{\nu} - s_{\mu}s_{\nu}$.

 {\em Definitions}  \cite{Hawking:1973uf,Wang:2003bt,Wang:2003xa}: A spatial 2-surface ${\cal{S}}$
  is said untrapped, marginally trapped, trapped, or anti-trapped according to
  \bq
  \lb{eq2.14}
   \left.\Theta_{+} \Theta_{-}\right|_{\cal{S}} = \begin{cases}
   < 0,  & \text{untrapped}, \cr
   = 0, & \text{marginally trapped},\cr
   > 0, \; \Theta_{\pm} < 0, & \text{trapped},\cr
    > 0, \; \Theta_{\pm} > 0, & \text{antitrapped}.\cr
   \end{cases}
  \eq
  {\em A black (white) hole horizon} is a marginally trapped that separates an untrapped region from a trapped one \cite{Hawking:1973uf,Wang:2003bt,Wang:2003xa}, while {\em a transition surface} is a marginally trapped that separates an anti-trapped region from a trapped one  \cite{Ashtekar:2018cay}.

  From Eqs.(\ref{eq2.13}) and (\ref{eq2.14}) we can see that  a black or white hole horizon does not exist, as now the BV spacetime is already geodesically complete, and no untrapped regions ($\Theta_{+} \Theta_{-} < 0$) in such a spacetime exist. However,
  transition surfaces could exist at $p_{c, \tau} = 0$, and such surfaces shall always separate   trapped ($p_{c, \tau} > 0$) regions from anti-trapped  ($p_{c, \tau} < 0$) ones. From Eq.(\ref{eq2.9}), on the other hand, it can be seen that this becomes possible when
   \bq
  \lb{eq2.15}
   \delta_c c(\tau) = n \pm \frac{\pi}{2},
  \eq
  where $n$ is an integer.  Numerically, we find that such surfaces indeed exist in the BV model. In fact, there exist infinite number of such surfaces. In particular, in Fig. \ref{fig3} (d) we show that two of such surfaces exist for $\tau \in (-3, 2)$.

\section{Conclusions and Remarks}

In this brief report, we adopted  the SS gauge \cite{Saini:2016vgo}, in which the lapse function is set to one, so the time-like coordinate becomes the cosmic time.  Then, we found that {\em black/white hole horizons do not exist}.  This conclusion is consistent with what we obtained previously by adopting a different gauge \cite{Gan:2022oiy}. This is  quite expected, as the physics should not depend on the gauge choice. The advantage of the SS gauge is that one can
easily show analytically that the BV spacetime is geodesically complete.

The above conclusion is important, as now the $\bar\mu$ scheme has been widely used in recent studies of LQBHs \cite{Han:2020uhb,Han:2022rsx}. Therefore, several comments now are in order. In particular, in  \cite{Han:2020uhb}
the authors considered the Lemaitre-Tolman-Bondi (LTB) spacetime
\bqn
\lb{eq3.1}
\text{d}s^2 = - \text{d}t^2 + \frac{\left(E^{\varphi}\right)^2}{|E^x|} \text{d}x^2 + |E^x| \text{d}^2\Omega,
\eqn
in which the Schwarzschild black hole solution is given by
\bqn
\lb{eq3.2}
E^x_{\text{GR}} &=& \left[\frac{3}{2}\sqrt{2m}(x - t)\right]^{4/3}, \nb\\
 E^{\varphi}_{\text{GR}} &=& \frac{2}{3} \left(\frac{3}{2}\sqrt{2m}\right)^{4/3}(x - t)^{1/3}.
\eqn
Note that the advantage of writing the Schwarzschild black hole solution in the LTB form is that it covers both inside and outside regions of the black hole. In particular,
the spacetime singularity now locates at $x - t = 0$, while the black hole horizon at $x - t =4m/3$.  Denoting the moment conjugates of $E^x$ and $E^{\varphi}$ by $K_x$ and $K_{\varphi}$, respectively, Han and Liu
considered the following replacements \cite{Han:2020uhb}
\bqn
\lb{eq3.3}
K_x \rightarrow \frac{\sin(\delta_x K_x)}{\delta_x}, \quad K_{\varphi} \rightarrow \frac{\sin(\delta_{\varphi} K_{\varphi})}{\delta_{\varphi}}.
\eqn
where
\bqn
\lb{eq3.4}
\delta_x = \frac{2\gamma \sqrt{\Delta |E^x|}}{E^{\varphi}}, \quad  \delta_{\varphi} = \frac{\gamma \sqrt{\Delta}}{\sqrt{|E^x|}}.
\eqn
Clearly, near the singularity, we have  $(\delta_x, \delta_{\varphi}) \rightarrow (0, \infty)$. Then, it is expected that quantum gravitational effects become very large, so in the reality the singularity used to appear classically now is smoothed out by these quantum effects, and a non-singular transition surface finally replaces the singularity. On the other hand, near the location of the classical black hole horizon, we have $(\delta_x, \delta_{\varphi})  \simeq \gamma \sqrt{\Delta}(2, (2m)^{-1})$, which are all
finite. Yet, for massive black holes, they are all very small, so  quantum effects near the horizons of these massive black holes are expected to be negligible. These are consistent with the results obtained in \cite{Han:2020uhb}.
Similar considerations were also carried out in \cite{Han:2022rsx}, so we expect that in this model black/white hole horizons also exist, and quantum effects near these horizons of massive black holes are expected to be negligible, too.

\begin{acknowledgments}
 The numerical computations were performed at the public computing service platform
 provided by TianHe-2 through the Institute for Theoretical Physics \& Cosmology,  Zhejiang University of Technology.
  YG  is partially supported by  the National Key Research and Development Program of China under Grant No. 2020YFC2201504.
  AW is partially supported by a NSF grant with the grant number: PHY2308845.

\end{acknowledgments}


%

\end{document}